\documentclass[traditabstract]{aa}
\usepackage{natbib}
\bibpunct{(}{)}{;}{a}{}{,}
\usepackage{graphicx}
\usepackage{revsymb}	
\usepackage{amsmath}	
\usepackage[usenames]{color}	
\usepackage{epstopdf}	
\newcommand{\ds}{\displaystyle}

\newcommand{\deriv} [2] {\frac {\textrm{d} #1 } {\textrm{d} #2} }

\newcommand{\eq}[1] {Eq.\,(\ref{#1})}

\usepackage{txfonts}

\newcommand{\numax}{\nu_{\mathrm{max}}}
\newcommand{\nuc}{\nu_{\mathrm{c}}}
\newcommand{\cs}{c_{\mathrm{s}}}

\begin{document}
\title{The underlying physical meaning of the $\numax-\nuc$ relation}

\author{K. Belkacem\inst{1,2} 
 \and M.J. Goupil\inst{3} \and M.A. Dupret\inst{2} \and R. Samadi\inst{3} \and F. Baudin\inst{1} \and A. Noels\inst{2} \and B. Mosser\inst{3}}

\institute{
Institut d'Astrophysique Spatiale, CNRS, Universit\'e Paris XI,
   91405 Orsay Cedex, France
\and
Institut dÕAstrophysique et de G\'eophysique, Universit\'e de Li\`ege, All\'ee du 6 Ao\^ut 17-B 4000 Li\`ege, Belgium 
\and 
 LESIA, UMR8109, Universit\'e Pierre et Marie Curie, Universit\'e
	Denis Diderot, Obs. de Paris, 92195 Meudon Cedex, France
}

   \offprints{K. Belkacem}
   \mail{Kevin.Belkacem@ulg.ac.be}
   \date{\today}

  \authorrunning{Belkacem et al.}
  \titlerunning{The underlying physical meaning of the $\numax-\nuc$ relation}

   \abstract{Asteroseismology of stars that exhibit solar-like oscillations are enjoying a growing interest 
  with the wealth of  observational results obtained with the CoRoT and Kepler missions. 
  In this framework, scaling laws between asteroseismic quantities  and stellar  parameters are becoming essential tools to study a rich variety of stars. However, the physical underlying mechanisms  of those scaling laws are still poorly known. Our objective is to provide a theoretical basis for the scaling  
   between the frequency of the maximum in the power spectrum ($\numax$) of solar-like oscillations 
   and the cut-off frequency ($\nuc$). 
   Using the SoHO GOLF observations together with theoretical considerations,
    we first confirm that the maximum of the height in oscillation power spectrum is determined
     by the so-called \emph{plateau} of the damping rates. 
     The physical origin of the plateau can be traced
     to the destabilizing effect  of the Lagrangian perturbation of entropy in the upper-most layers which
     becomes important when the modal period and the local thermal relaxation time-scale are comparable. 
     Based on this analysis, we then find a linear relation between $\numax$ and $\nuc$, with a 
     coefficient that depends on the ratio of the Mach number of the exciting turbulence to the third power to the mixing-length parameter.}

   \keywords{Convection - Turbulence - Stars: oscillations - Stars: interiors}

   \maketitle
   
\section{Introduction}
\label{intro}

 Scaling relations  between 
 asteroseismic quantities  and stellar  parameters such as stellar mass, radius, effective temperature and
  luminosity  have been observationally  derived  by several authors
  \citep[e.g.][]{Kjeldsen95,Chaplin08,Chaplin09,Stello09b} using ground-based data.
   More recently, the space-missions CoRoT and Kepler confirmed those
    results by providing accurate and homogeneous measurements  
    for a large sample of stars from red giants to main-sequence stars \citep[e.g., ][]{Mosser10}.  
Scaling relations are essential to study 
 a large set of stars \citep[e.g., ][]{Kallinger09,Stello09a}  for which, in general, little is known, 
 to provide a first order estimate for mass and radius \citep[e.g., ][]{Basu09,Mosser10}, or  
 to probe the populations of red giants \citep{Miglio09}.
  
  Scaling laws can also lead to a better understanding of  the underlying physical mechanisms
   governing the energetical behaviour of modes.
In particular, it has been conjectured by \cite{Brown91} that the frequency of
 the maximum of the power spectrum ($\numax$) scales as 
 the cut-off frequency $\nuc$ 
  because  the latter corresponds to a typical time-scale of the atmosphere.
   The continuous increase of detected stars with solar-like oscillations 
   has then  confirmed this relation  \cite[e.g., ][]{Bedding03,Stello09b}. 
However, the underlying physical origin of this scaling relation is still poorly understood. 
Indeed, $\numax$ is associated with the coupling between 
turbulent convection and oscillations and results from a balance between the damping   
and  the driving of the modes. The cut-off frequency is associated with  
the mean surface properties of the star and the sound speed, making the origin of 
the $\numax - \nuc$ relation very intriguing. 

As a first step toward an understanding, one has to determine which of the  damping rate  
or the excitation rate  is the main  responsible for the maximum of power in the observed spectra. 
\cite{Chaplin08}, using a theoretical approach, pointed out that in the solar case  $\numax$ coincides 
with the plateau of the linewidth variation with frequency. 
We will confirm this result using observations from the GOLF instrument in the solar case. 
However, several issues remain to be addressed: 
is $\numax$ for any star directly 
related with  the observed  plateau in  the mode-widths variation with frequency? 
In case of a positive answer, what is the origin of this  relation? 
The first issue is quite difficult to answer  as it is expected to strongly 
depend on the model used for the description of the pulsation-convection interaction. 
 Nevertheless, CoRoT observations begin to answer this issue and several stars (HD49933, HD180420, HD49385, and HD52265) suggest that $\numax$ correponds to the plateau of the damping rates \cite[see][for details]{Benomar09b,Barban09,Deheuvels09,Ballot11}. 
The second step consists in determining the main physical causes responsible
 for the plateau of the damping rates and its mean frequency ($\nu_{\rm \Gamma}$). 
 Subsequently, one has to determine a general scaling law that relates 
 the frequency of the plateau of the damping rates to the stellar parameters. 
In this paper,  we discuss the first issue then we focus on 
  the second issue by deriving a  theoretical  relation 
  between $\nu_{\rm \Gamma}$ and $\nuc$.  
  If one accepts the positive answer to the first issue,  
  this also provides the scaling relation between $\numax$ and $\nuc$. 
  
This paper is organised as follows. In Sect.~2 we present the observed 
scaling law obtained from a homogeneous set of CoRoT data and 
  show that the maximum mode height in the solar power spectrum 
  coincides with a marked minimum of the mode-width when corrected from mode inertia. 
  We then point out, in Sect.~3, that such a minimum is the result of a destabilizing effect in the super-adiabatic region.  The relation between $\nu_{\rm \Gamma}$ and $\nuc$ is  demonstrated in Sect. 4, and  conclusions are provided in Sect.~5. 

\section{The observed scaling law}
\label{observations}

We use the CoRoT seimological field data to ensure a homogeneous sample: HD49933 \citep{Benomar09b}, HD181420 \citep{Barban09}, HD49385 \citep{Deheuvels09}. 
We also use the results on HD50890 \citep{Baudin11} and on HD181907 \citep{Carrier09}, a red giant, and the Sun. The characteristics of these stars are listed in Table\,\ref{tab:HD}, as well as the way their fundamental parameter is obtained. 

For the Sun, due to the presence of pseudo-modes above the cut-off frequency \cite[e.g. ][]{garcia98}, 
observational determination of $\nu_{\rm c}$ is not obvious. Nevertheless, one can infer 
 a theoretical relation for this frequency 
$\omega_{\rm c} = \cs / 2 H_\rho \propto g / \sqrt{T_{\rm eff}} \propto M\, R^{-2}\, T_{\rm eff}^{-1/2}$
\citep[e.g., ][]{Balmforth90}, where $\cs$ is the sound speed, $H_\rho$ the density scale height, 
$g$ the gravitational field, $M$ the mass, $R$ the radius,  and $T_{\rm eff}$ the temperature at the photosphere. 
 When scaled to the solar case, this relations becomes 
\begin{align}
\label{scaling}
\nu_{\rm c} = \nu_{\rm c \odot} \left(\frac{M}{M_\odot}\right) \, \left(\frac{R}{R_\odot}\right)^{-2} \, \left(\frac{T_{\rm eff}}{T_{\rm eff \odot}}\right)^{-1/2}  \, , 
\end{align}
with $\nu_{\rm c \odot} = 5.3 $mHz, and $M_\odot, R_\odot, T_{\rm eff \odot}$ the solar values of mass, radius, and effective temperature respectively. Note that we will assume $H_\rho = H_{\rm p} = P /\rho g$ with $P,\rho$  respectively denoting pressure,
 density. This is a commonly used approximation \citep[e.g., ][]{Stello09b} that presupposes an isothermal atmosphere, which is of sufficient accuracy for our purposes. 

\begin{figure}
\begin{center}
\includegraphics[height=9cm,width=6cm,angle=90]{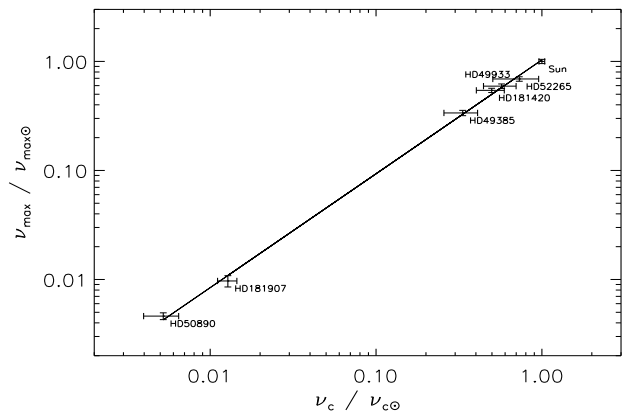}
\caption{ Frequency of the maximum of oscillation power for the main- 
sequence and red-giant stars of Table~\ref{tab:HD} as a function of the frequency cut-off. 
All quantities are normalized to the solar values.}
\label{results}
\end{center}
\end{figure}

Using the stars listed in Table~\ref{tab:HD}, 
 and their measured $\nu_{\rm max}$, the relation between $\nu_{\rm max}$ 
 and $\nu_{\rm c}$ is displayed in Fig.~\ref{results}. 
It relies on two kinds of results: direct observations of $\nu_{\rm max}$ in the spectrum of the star on one hand, and on estimates of the mass ($M$), radius ($R$) and effective temperature ($T_{\rm eff}$) of the star on the other hand. The latter are derived from photometric or spectroscopic observations but can be derived in some cases from stellar modelling. Here, $M$ and $R$ must be derived from stellar modelling and not from scaling laws as the aim of this work is to establish such a scaling law. 
 The strict proportionality (the fitted slope is $1.01\pm 0.02$) is clearly seen from this sample spanning from the Sun to a luminous red giant (HD50890). 
This is in agreement with the results obtained by several authors for main-sequence stars \citep[e.g.,][]{Bedding03}, as well as red giants \citep[e.g.][]{Mosser10}. 
The issue is now to assess the physical background underlying this relation. 

\begin{table}[]
\center{
\begin{tabular}{|c|c|c|c|c|}
\hline
Star name & $T_{eff}$ (K) & $M/M_{\odot}$ & $R/R_\odot$ & $\nu_{\rm max}$ ($\mu$Hz) \\
\hline
Sun & 5780 & 1 & 1 & 3034 \\
HD49933 & 6650 & 1.2 & 1.4 & 1800 \\
HD181420 & 6580 & 1.4 & 1.6 & 1647 \\
HD49385 & 6095 & 1.3 & 1.9 & 1022 \\
HD52265 & 6115 & 1.2 & 1.3 & 2095 \\
HD181907 & 4760 & 1.7 & 12.2 & 29.4 \\
HD50890 & 4665 & 4.5 & 31 & 14 \\
\hline
\end{tabular}}
\caption{Stellar characteristics (from the literature - see references in Sect~\ref{observations}) for the stars used in the comparison with the present results. For HD49933, $\nu_{\rm max}$ and $T_{\rm eff}$ are taken from \citet{Benomar09b}; $M$ and $R$ are taken from \citet{Benomar10}. For HD181420, $\nu_{\rm max}$ and $T_{\rm eff}$ are taken from \citet{Barban09}; $M$ and $R$ are provided by M.-J. Goupil (private communication). For HD49385, $\nu_{\rm max}$ and $T_{\rm eff}$ are taken from \citet{Deheuvels09}; $M$ and $R$ are provided by M.-J. Goupil (private communication). For HD181907, $\nu_{\rm max}$ and $T_{\rm eff}$ are taken from \citet{Carrier09}. For HD50890, $\nu_{\rm max}$, $T_{\rm eff}$, $M$ and $R$ are taken from \citet{Baudin11}.
}
\label{tab:HD}
\end{table}

\section{Height maximum in the power spectrum}

In this section, we confirm  that the maximum of the power spectrum of solar-like 
oscillations is related to the plateau of the line-width by using solar observations 
from the GOLF instrument and we then discuss the physical origin of 
the depression of the damping rates (i.e. the plateau). 

\subsection{Origin of the maximum of height in the power density spectrum} 
\label{Max_H}

We consider the height $H$ of a given mode in the power spectrum, which is a natural observable. 
To derive it, let us first define the damping rate of the modes given by \citep[e.g., ][]{Dupret09}
\begin{equation}
\label{workeq}
\eta = \frac{-W}{2\,\omega \, |\xi_r(R)|^2 \mathcal{M}}\,,
\end{equation}  
where $\omega$ is the angular frequency, 
$W$ is the total work performed by the gas during one oscillation cycle, 
$\vec{\xi}$ is the displacement vector, and 
$\mathcal{M}$ is the mode mass
\begin{equation}
\label{ineq}
 \mathcal{M}=   \int_0^M \frac{|\vec{\xi}|^2}{|\xi_r(R)|^2} \,{\rm d}m  \, .
\end{equation}
$\xi_r(R)$ corresponds to the radial displacement at the layer where the oscillations
are measured, {\bf $M$ is the total mass of the star. } 

For stochastically excited modes, 
the power injected into the modes is \citep[e.g., ][]{Samadi00I,Belkacem06b}
\begin{equation}
\label{stocheq}
P\;=\;\frac{1}{8\, \mathcal{M}}\,(C_R^2+C_S^2)\,,
\end{equation}
where $C_R^2$ and $C_S^2$ are the turbulent Reynolds stress and entropy contributions,
respectively.
We then introduce the height of the mode profile in the power spectrum, which is an observable, as   
\citep[see e.g.][]{Chaplin05,Belkacem06b} 
\begin{equation}
\label{heighteqres}
H\;=\;\frac{P\, }{2\,\eta^2\,\mathcal{M}} \, .
\end{equation}

However, it is useful to express $H$ in a form that does not explicitly depend
 on the mode mass ($\mathcal{M}$). To this end, we note from Eqs.~(\ref{workeq}) and (\ref{stocheq}) that both the excitation $P$ and the damping rate $\eta$ are inversely proportional to the mode mass. 
Hence, to disentangle the effect of the driving and damping from the effect of mode mass, we introduce the quantities $\Pi= \mathcal{P} \, \mathcal{M}$ and $\Theta= \eta \, \mathcal{M}$, independent of mode masses. Then, using \eq{heighteqres}, the expression of the mode height becomes 
\begin{align}
\label{height}
\ds H = \frac{\Pi}{2 \Theta^2}  \, . 
\end{align}

\begin{figure}
\begin{center}
\includegraphics[height=6cm,width=9cm]{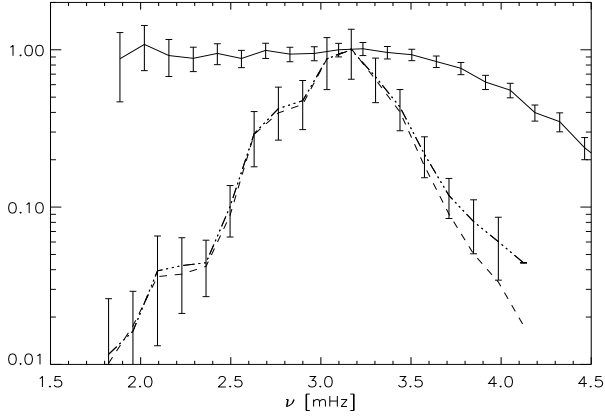}
\caption{Normalized $1/ (2 \Theta^2)$ (dashed-dots line) , where $\Theta=\eta \, \mathcal{M}$, and $\Pi=\mathcal{P\, M}$ (solid line) 
computed from solar seismic data from  the GOLF instrument \citep{Baudin05}. The normalized mode height ($H$) corresponds to the dashed line. 
All quantities are normalized to unity at the frequency of $\nu=\nu_{\rm max}$. 
The associated mode masses are computed as described in \cite{Belkacem06b}.} 
\label{fig:golf:PM}
\end{center}
\end{figure}

Figure~\ref{fig:golf:PM} displays 
the variations of  $H$ with mode frequency as well as its two contributions $\Pi$ and 
$1/\Theta^2$. One can clearly distinguish a maximum for $H$ near $\nu \simeq 3.2$ mHz 
that corresponds to the $\nu_{\rm max}$ frequency. $\Pi$ remains roughly constant (efficient driving  regime)  
except at high frequency  beyond $\nu_{\rm max}$  which corresponds to the inefficient driving regime
  \citep[see][for details]{Samadi00I}. 
 On the other hand  $1/\Theta^2$   shows a sharp maximum and its variation clearly  dominates over that of
 $\Pi$ and controls the variation of $H$ and the apparition of its maximum.
 We conclude that the maximum of $H$ is determined by the minimum of $\Theta^2$ and corresponds to the plateau of the line widths.
 In other words, the depression (plateau) of the damping rates $\eta$ 
 is responsible for the presence of a maximum in the power spectrum, in agreement with \cite{Chaplin08}.  

\subsection{Origin of the depression of the damping rates}
\label{depression}

\cite{B92a} mentioned  that the depression of the solar damping rates originate from a
 destabilising effect in the super-adiabatic layer. He also stressed that the
  plateau of the damping rates occurs when there is a resonance between the thermal time 
  scale and the modal frequency. 

Following these ideas, we  use  the MAD non-adiabatic pulsation code \citep{MAD02} for computing the solar  damping rates.  
This code includes a time-dependent convection treatment \citep{MAD05}   different from that by Balmforth
(1992). Nevertheless, we reach  the same conclusion (see Appendix~\ref{app_B} for details):
the responsible for the destabilizing effect is the Lagrangian perturbation 
of entropy ($\delta S$) that exhibits a rapid variation mainly 
 in the super-adiabatic layer as well as in the atmospheric layers (see Appendix~\ref{origin_depression} and Fig.~\ref{work_theo}). 

To understand the origin of such an oscillation and illustrate 
the occurrence of the resonance, we consider the super-adiabatic layers
 and we examine the case of a highly non-adiabatic solution 
 \citep[see][ for the case of a purely radiative envelope]{Pesnell84}. 
 We assume that Lagrangian perturbations of radiative and convective
  luminosities are dominated by perturbations of entropy (see \eq{deltaLr3} and \eq{approx_Lc}). 
  This leads to a second-order equation for the {\bf entropy perturbations $\delta S$} (\eq{eq_osc_S_second_order}, see Appendix.~\ref{oscill_s} for the  derivation). 
  To obtain a more explicit solution for $\delta S$, we further employ the dimensional approximation ${\rm d} \delta L / {\rm d}r \backsim  \delta L / H_p$, so that 
\begin{align}
\label{equation_osc}
\deriv{}{\ln T} \left( \frac{\delta S}{c_v} \right)+ \lambda \left( \frac{\delta S}{c_v} \right) = 0 \,, \quad {\rm with} \quad 
\lambda = \mathcal{A} - i \, \mathcal{B}\, , 
\end{align}
where $c_v = \left(\partial U / \partial T \right)_\rho$ with $U$ the internal energy, $\mathcal{A}$ and $\mathcal{B}$ are defined by 
\begin{align}
\label{lambda}
\mathcal{A} &=  \left(\frac{L_c}{L} \psi \deriv{\ln c_v}{\ln T} + \frac{L_R}{L} (4-\kappa_T) \right)\left( 1 + (\psi-1) \frac{L_c}{L} \right)^{-1}  \nonumber \\
\mathcal{B} &= \mathcal{Q} \left[ 1 + (\psi-1) \frac{L_c}{L} \right]^{-1}  \, ,  
\end{align} 
where $\kappa_T = \left(\partial \ln \kappa / \partial \ln T \right)_\rho$, $L_c$, and $L_R$ are the convective and radiative luminosity respectively, $T$ the temperature, $\psi$ is defined by \eq{def_psi}, and we have defined the ratio $\mathcal{Q}$ such as
\begin{align}
\label{definition_Q}
\mathcal{Q} =\omega \tau \, , \quad \mbox{with} \quad \tau^{-1} = \frac{L}{4\pi r^2\rho c_v T H_p} = \tau_{\rm conv}^{-1} + \tau_{\rm rad}^{-1}
\end{align}
with $\omega=2 \pi \, \nu$, $\nu$ the modal frequency, $\tau$ a local thermal time-scale, $\tau_{\rm rad}$ and $\tau_{\rm conv}$ the radiative and convective thermal time-scales, respectively. 
From \eq{equation_osc}, the oscillatory part of the final solution is 
$\left( \delta S / c_v \right) \propto \exp \left[- {\rm i} \, \int \mathcal{B} \,  {\rm d}\ln T \right]$, 
which describes the oscillatory behaviour of entropy perturbations in the super-adiabatic  layers. 

As discussed in  Appendix B3 (Fig B2 top),  all modes in the range of interest  
 have a similar negative integrated work, $W$, at the bottom of the superadiabatic layer. This corresponds to a
 large damping at this level in the star. 
 In the superadiabatic layers, 
 the entropy's  oscillatory behaviour controls the oscillating behavior of  $W$. 
 When the pulsation period and thus the wave-length of the entropy perturbations
are too large ($ \mathcal{Q} \ll1$), 
  the destabilizing contribution has not grown enough; 
the cumulated work $W$ increases too slowly. The net result at the surface  is a large damping. 
  When the period is too small ($ \mathcal{Q}  \gg1$),  the  rapid oscillation of the entropy perturbation causes
  a rapid oscillation of $W$ which increases and again decreases before reaching the surface and the net
  result at the surface is again  a large damping.   Those two limits correspond to low and high frequencies, \emph{i.e.} to the two branches of $1/\Theta^2$ displayed in Fig.~\ref{fig:golf:PM}. 
  A minimum damping is then obtained  for a period neither too small nor too large i.e. $ \mathcal{Q} \simeq 1$ where the destabilizing contribution nearly but not quite compensates the strong damping of the  layers below the super adiabatic layers.
  
\begin{figure}[!]
\begin{center}
\includegraphics[height=6cm,width=9cm]{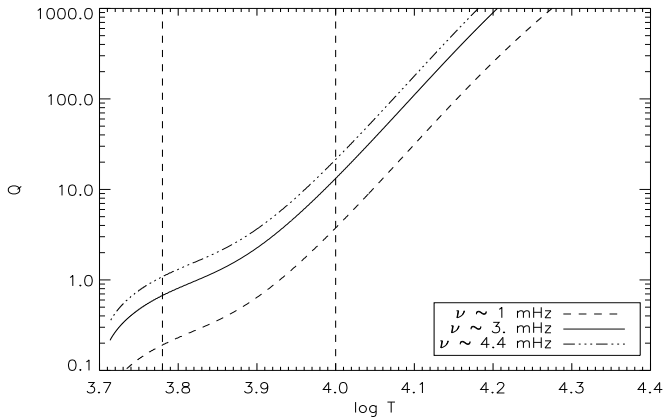}\\
\caption{Product $\mathcal{Q}$ (see \eq{definition_Q}) versus the logarithm of the temperature for three values of the mode frequency, for a solar model described in Appendix~\ref{app_B}. Vertical dotted lines delimit the limits of super-adiabatic gradient (see Fig.~\ref{work_theo} bottom panel).}
\label{figQ}
\end{center}
\end{figure}
 
The value of $\mathcal{Q}$ is illustrated  in Fig.~\ref{figQ} for three modes. It confirms  that the resonance  $\mathcal{Q} \simeq 1$ occurs in the super-adiabatic region for the mode with frequency  $\nu \simeq \nu_{\rm max}$.
Hence from the $\mathcal{Q}$ definition \eq{definition_Q}, one derives the resonance condition 
\begin{align}
\label{resonance}
  \nu_{\rm max} \simeq \frac{1}{2\pi \, \tau} \, .
\end{align}

\section{Derivation of the scaling law}
\label{deriv_scaling}

We now turn to the relation between the thermal time-scale ($\tau$) and the cut-off frequency. 
 To this end, we use a grid of stellar models for masses between $M=1 \, M_\odot$ and $M=1.4 \,M_\odot$ from the ZAMS to the ascending {\bf vertical} branch, typical of observed solar-like pulsators. 
The grid is obtained by using the  stellar evolution code CESAM2k  \citep{Morel97,Morel08}. The atmosphere is computed assuming a grey Eddington atmosphere.
Convection is included according to  B\"{o}hm-Vitense mixing-length (MLT) formalism. The mixing-length parameter is $\alpha=1.6$.  The chemical composition follows \cite{Asplund05}, with an helium mass fraction of $0.2485$. All quantities are evaluated at the maximum of the super-adibatic gradient, which corresponds to the maximum of $\delta S$ (see Sect.~\ref{depression}) and the location of the resonance (see \eq{resonance}). 

\begin{figure}[!]
\begin{center}
\includegraphics[height=6cm,width=9cm]{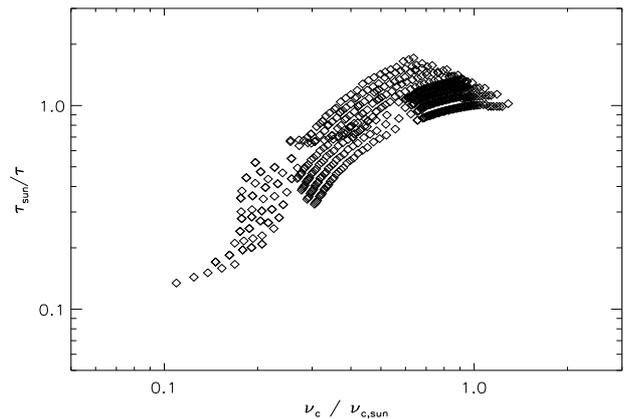}\\
\includegraphics[height=6cm,width=9cm]{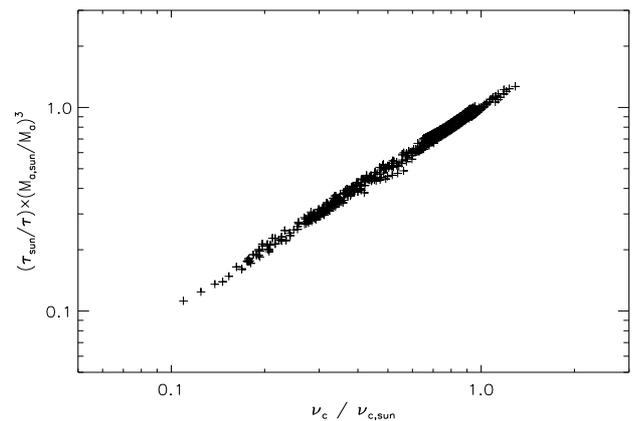}\\
\caption{{\bf Top:} Thermal frequency ($1/\tau$) computed from \eq{definition_Q} versus the cut-off frequency computed following \eq{scaling}, normalized to the solar values, for models with masses ranging from $M=1 M_\odot$ to $M=1.4 M_\odot$ (with an increment of $0.05 M_\odot$) and from the ZAMS to the ascending vertical  branch. 
{\bf Bottom:} The same as for the top panel, except the thermal frequency is divided by the Mach number to the third versus the normalized cut-off frequency.}
\label{fig_grille}
\end{center}
\end{figure}

From Fig.~\ref{fig_grille} (top), the relation between the thermal frequency ($1/\tau$) and the cut-off frequency ($\nu_{\rm c}$) is close to linear but still shows a significant dispersion. 
More precisely, the relation between those two frequencies is approximatively linear and the dispersion is related to the dispersion in mass, in agreement with observations \citep[e.g. ][]{Mosser10}. 
We then conclude that the observed relation between $\nu_{\rm max}$ and $\nu_{\rm c}$ is in fact the result of the resonance between $\nu_{\rm max}$ and $1/\tau$, as well as the relation between $1/\tau$ and $\nu_{\rm c}$. 

To go further, let us investigate the relation between $1/\tau$ and $\nu_{\rm c}$. First, \eq{definition_Q} can be recast as 
\begin{align}
\label{def_thermal}
\frac{1}{\tau} = \frac{F_{\rm conv}}{\rho c_v T H_p}  \left[ 1 + \frac{F_{\rm rad}}{F_{\rm conv}} \right] \, , 
\end{align}
where $F_{\rm conv}$ and $F_{\rm rad}$ are the convective and radiative fluxes, respectively. 
The MLT solution for the convective flux and the convective rms velocity can be written \citep[see][ for details]{CoxGiuli68}
\begin{align}
\label{Fc}
F_{\rm conv} &= \frac{1}{2} \rho c_p \texttt{v}_{\rm conv} T \frac{\Lambda}{H_p} \left(\nabla - \nabla^\prime\right) \\
\label{Vc}
\texttt{v}_{\rm conv} &= \frac{\alpha \cs \Sigma^{1/2}}{2\sqrt{2} \Gamma_1^{1/2}} \left(\nabla - \nabla^\prime\right)^{1/2}
\end{align}
where $\Lambda=\alpha H_p$ is the mixing length, $\alpha$ the mixing-length parameter, $\nabla=\left( {\rm d} \ln T / {\rm d} \ln P \right)$, $\nabla^\prime=\left( {\rm d} \ln T^\prime / {\rm d} \ln P \right)$ the gradient of rising convective element, 
$\Sigma= \left(\partial \ln \rho/ \partial \ln T\right)_{\mu,P}$, with $\mu$ the mean molecular weight, and $\Gamma_1=\left(\partial \ln P / \partial \ln \rho \right)_{\rm ad}$. 
Now, by inserting \eq{Fc} and \eq{Vc} into \eq{def_thermal}, one obtains 
\begin{align}
\label{thermal_final}
\frac{1}{\tau} = 8 \left(\frac{\Gamma_1^2}{\chi_\rho \Sigma} \right) \, \left(\frac{\mathcal{M}_a^3}{\alpha}\right) \, \left(\frac{\cs}{2 H_p} \right) \left[ 1 + \frac{F_{\rm rad}}{F_{\rm conv}} \right] 
\end{align}
where $\mathcal{M}_a=\texttt{v}_{\rm conv} / \cs$ the Mach number, and $\chi_\rho = \left(\partial \ln P/ \partial \ln \rho \right)_T$. 

We verified that for a given physic, the ratio $F_{\rm rad} / F_{\rm conv}$ is approximately the same for all the models considered in the super-adiabatic layer. 
Hence, by use of \eq{thermal_final} as well as the resonance condition (\eq{resonance}), we conclude that 
\begin{align}
\label{thermal_approx}
\nu_{\rm max} \propto \frac{1}{\tau} \propto \left(\frac{\Gamma_1^2}{\chi_\rho \Sigma} \right) \, \left(\frac{\mathcal{M}_a^3}{\alpha}\right) \, \nu_{\rm c} \, , 
\end{align}
which is the observed scaling between $\nu_{\rm max}$ and $\nu_{\rm c}$ (see Fig.~\ref{results}), since the thermodynamic quantities hardly vary.  

Equation~(\ref{thermal_approx}) describes the observed scaling between $\nu_{\rm max}$ and $\nu_{\rm c}$ (see Fig.~\ref{results}) but also shows that most of the departure from the linear relationship between $1/\tau$ and $\nu_{\rm c}$ comes from the Mach number, as confirmed by Fig.~\ref{fig_grille} (bottom panel). We also point out  that as shown by Fig.~\ref{fig_grille} and Fig.~\ref{results} for the main-sequence stars, the departure from the linear relationship is of the same order of magnitude as the uncertainties on the cut-off frequency. However, our grid of models is not suited for a proper comparison between the observations and the theoretical relation. This work is definitely desirable in the future. 

\section{Conclusion}

We have addressed the issue of the physical reason for  the existence
 of a scaling relation between $\nu_{\rm max}$ and $\nu_{\rm c}$. 
 We have found that the depression of the damping rates determines 
 $\nu_{\rm max}$ because there is a resonance between the local thermal time-scale
  in the super-adiabatic region and the modal period. 
 This implies  that $\nu_{\rm max}$ does not scale 
   only with $\nu_{\rm c}$ but also with the ratio $\mathcal{M}_a^3 / \alpha$. 
   As pointed out in Sect.~\ref{intro}, 
   the observed scaling between $\nu_{\rm max}$ and $\nu_{\rm c}$ 
   is not obvious at first glance since the first frequency depends on the dynamical 
   properties of the convective region while the second is a static
     property of the surface layers. The additional dependence 
     the Mach number resolves this paradox. 

This scaling relation is potentially a powerful probe to constraint 
the dynamical properties of the upper-most layers of solar-like pulsators 
through the ratio $\mathcal{M}_a^3 / \alpha$. Indeed, as shown in this paper, 
most of the dispersion in the $\nu_{\rm max}-\nu_{\rm c}$ scaling is related to 
the Mach number. The investigation of the ratio between $\nu_{\rm max}$ and $\nu_{\rm c}$ 
in main-sequence stars, subgiants, and red giants may give us statistical information on the evolution 
of the properties of turbulent convection from main-sequence to red giant stars, 
through for instance the mixing-length parameter. Indeed, a future work will consist in computing 
models, that correspond to the observations, and to make a comparison between the observed and theoretical dispersion from the linear relation between $\nu_{\rm max}$ and $\nu_{\rm c}$. 

In other specific cases, for which stellar parameters are well known 
(e.g., in pulsating binaries) the relation between $\nu_{\rm max}$ and $\nu_{\rm c}$  
could gives us directly the value of the Mach number in the upper-most convective layers. 

\begin{acknowledgements}

K. B. gratefully acknowledges support from the CNES (ÒCentre National dÕEtudes
SpatialesÓ) through a postdoctoral fellowship. 

\end{acknowledgements}


\newpage
\appendix

\section{The \emph{plateau} of the damping rates} 
\label{app_B}

\subsection{Computation of the damping rates}
\label{compute_damping}

\begin{figure}[!]
\begin{center}
\includegraphics[height=6cm,width=9cm]{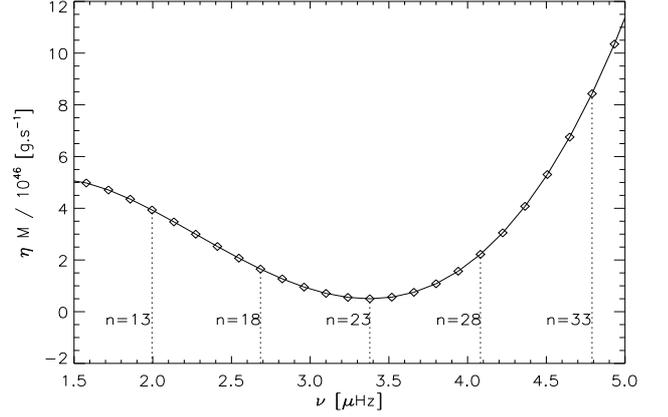}\
\caption{Product of the damping rates ($\eta$) times the mode mass ($\mathcal{M}$) versus mode frequency. The vertical dotted lines identify radial mode orders.}
\label{eta_theo}
\end{center}
\end{figure}

Damping rates have been computed with the non-adiabatic 
pulsation code MAD \citep{MAD02}. This code includes a time-dependent convection 
(TDC) treatment described in \cite{MAD05}. This formulation involves   
a free parameter $\beta$ which takes complex values  and enters the
perturbed  energy equation. This parameter was introduced 
to prevent the occurence of  non-physical spatial oscillations in the eigenfunctions.  
We use here the value $\beta=-0.55-1.7 i$ which is calibrated so  that resulting 
 damping rates   reproduce 
the variation of  the solar damping rates $\eta$ with frequency and more precisely 
 the depression of the $\eta$ profile (see Fig.~\ref{eta_theo}). 
Note that TDC  is a local  formulation of  convection.  
This simplifies the theoretical  description and is sufficient 
here as we seek for   a qualitative understanding 
of the relation between the frequency location of the damping rate 
depression and the cut-off frequency. We stress that the above  approximations 
do not influence qualitatively the conclusions. 

This approach takes into account the role played 
by the variations of the convective flux, the turbulent pressure, and the dissipation rate of 
turbulent kinetic energy. Hence, the integral expression of the damping can be written as follows
\begin{equation}
\label{dampings_radiatif}
\eta = \frac{1}{2 \, \omega \mathcal{M} \, \vert \xi_r (R)\vert^2} \int_{0}^{M} \mathcal{I}m \left[
\frac{\delta \rho}{\rho}^*  \left( \frac{\delta P_{\rm turb}}{\rho} + 
\left(\Gamma_3 - 1\right)  T \delta S 
\right) \right] \textrm{d}m
\end{equation}
where $\xi_r (R)$ is the radial mode displacement at the photosphere, $\omega$ the mode frequency, 
$\rho$ the mean density, 
$\Gamma_3-1 = \left( \partial \ln  T/ \partial \ln \rho \right)_S$, $T$ the unperturbed
temperature  and the star denotes the complex conjugate. The symbol $\delta$ represents a Lagrangean
perturbation: 
$\delta S$ is the perturbation of specific entropy, 
$\delta \rho$ the density perturbation, 
$\delta P_{\rm turb}$ the perturbation of turbulent pressure. 
The quantity $\delta P_{\rm turb} / \rho$ represents the contribution of turbulent pressure while the second term $(\Gamma_3-1) \, T\, \delta S$ includes the variations of radiative and convective fluxes as well as  the dissipation rate of turbulent kinetic energy, as given by the energy conservation equation
\begin{align}
\label{energy}
i \sigma T \delta S = - \deriv{\delta L_r}{m} - \deriv{\delta L_c}{m} + \delta \epsilon_t
\end{align}
with $\delta L_r, \delta L_c$ being the perturbations of the radiative and convective fluxes respectively,  $\delta \epsilon_t$ the perturbation of the dissipation rate of turbulent kinetic energy into heat, and $\sigma = \omega + i \eta$. Note that \eq{energy} is only valid for radial modes we are interested in. 

\subsection{Origin of the depression of the damping rates}
\label{origin_depression}

\begin{figure}[!]
\begin{center}
\includegraphics[height=6cm,width=9cm]{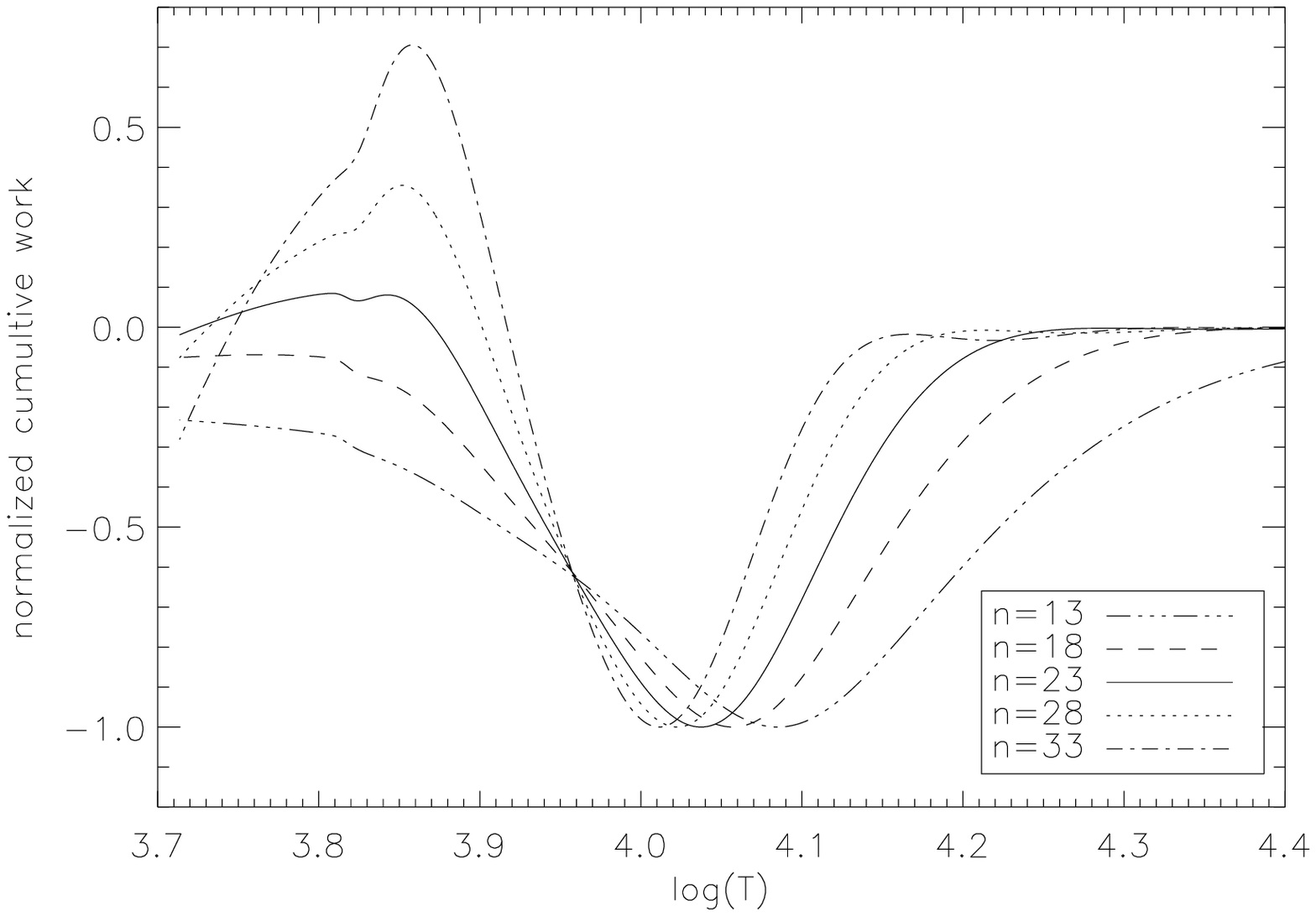}\\
\includegraphics[height=6cm,width=9cm]{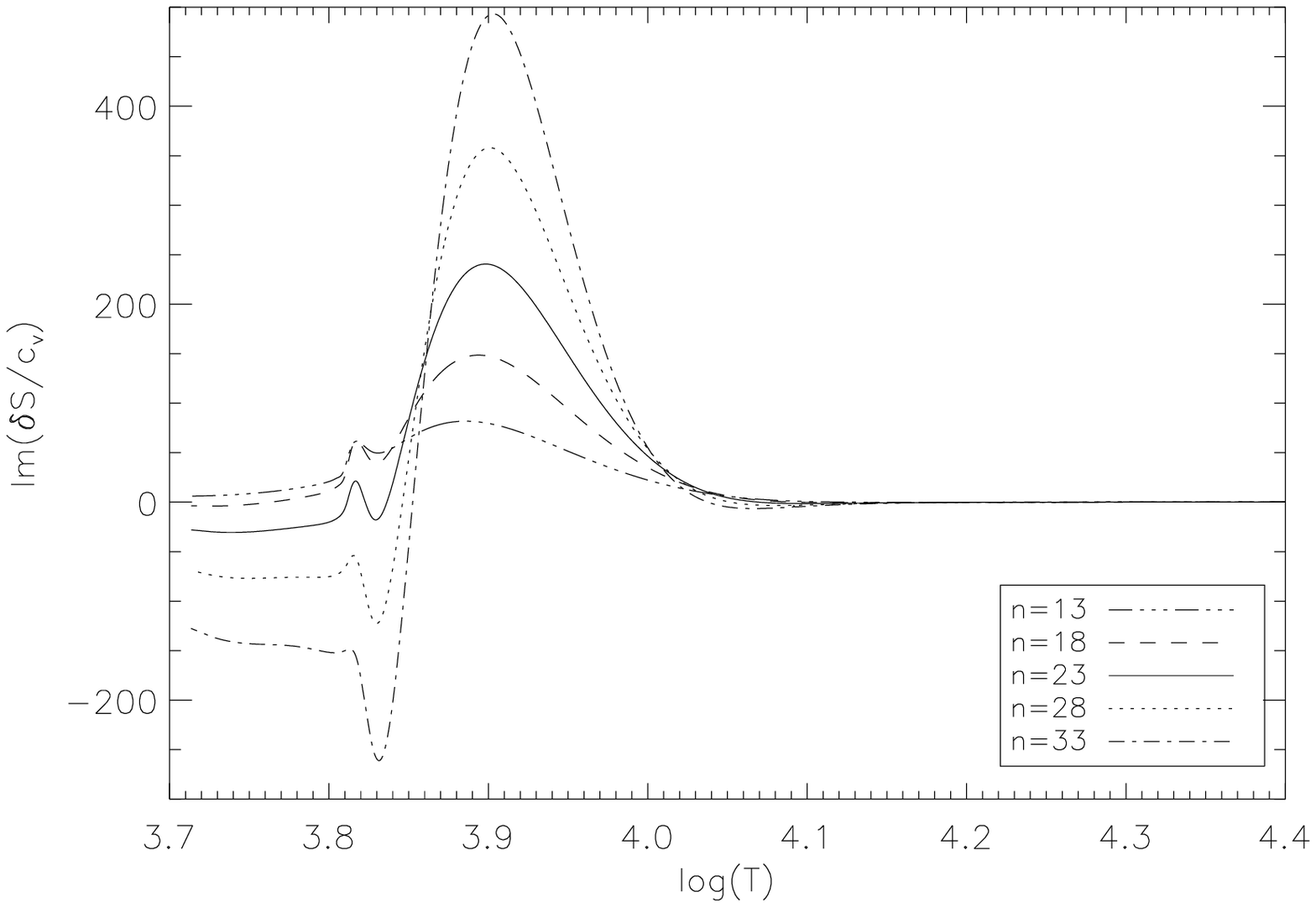}\\
\includegraphics[height=6cm,width=9cm]{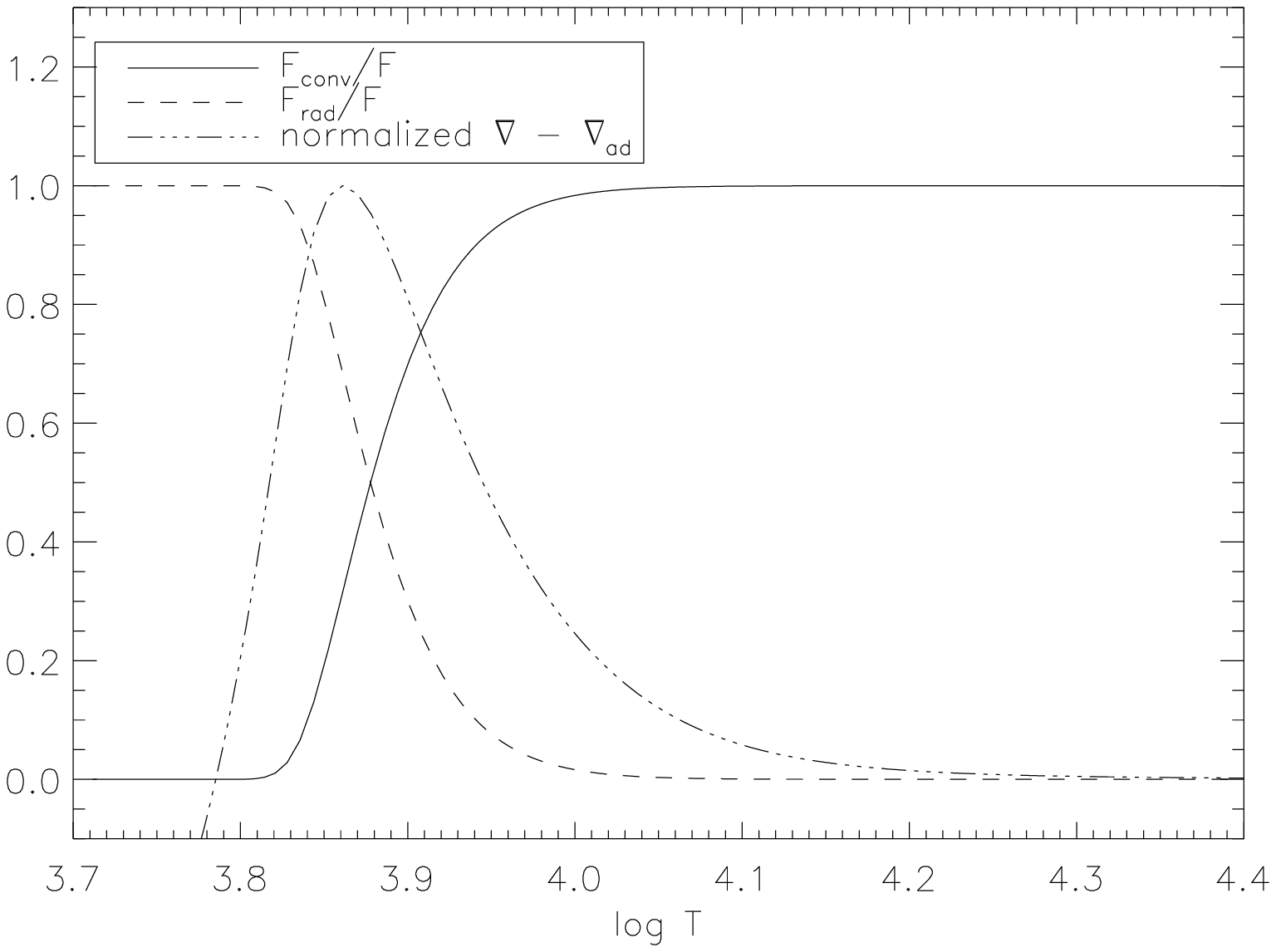}
\caption{{\bf Top:} Normalized cumulated work integral versus 
logarithm of temperature for five values of eigenfrequencies. 
These modes are emphasized in Fig.~\ref{eta_theo}. 
{\bf Middle:} Imaginary part of the Lagrangian perturbation of entropy versus  logarithm of temperature. 
{\bf Bottom:} Normalized convective and radiative fluxes versus logarithm of temperature. 
The difference between the real and adiabatic gradient ($\nabla$) is also plotted and normalized to its maximum.}
\label{work_theo}
\end{center}
\end{figure}

The depression of the damping rates, located around $\nu \sim 3.5$ mHz (Fig.\ref{eta_theo}), 
results from a subtle balance between the  above contributions to the work integral. 
The cumulated work integral (regions where it increases outwards drive the oscillation 
and regions where it decreases outwards damp the oscillation) 
allows us to identify the processes that create this depression. 
Fig.~\ref{work_theo} (top) shows that mode damping results from  stabilizing effects 
from  inner layers at temperature greater than $\log T \backsim 4$ 
destabilizing effects in the upper layers located in the super-adiabatic layers 
(\emph{i.e.} between $\log T \backsim 3.95$ and $\log T \backsim 3.8$)  and  for  high  radial order modes 
again stabilizing effects from the  very outer layers.
Hence, the behavior of {\bf the product $\Theta$ of the damping rates to the mode mass}, 
which is the integral appearing in \eq{dampings_radiatif}, 
can then be described  as follows: for modes with frequencies $\nu \leq \nu_{\rm max}$
 the higher the mode frequency the larger 
the contribution of the destabilizing region  and  $\Theta$ keeps on decreasing.
  For $\nu > \nu_{\rm max}$, despite an increasing contribution of the superadiabatic boundary layers, 
  atmospheric layers stabilize the modes resulting in an increase of $\Theta$.
At $\nu=\nu_{\rm max}$, compensation is  maximal  giving rise to the  minimum of $\Theta$.

The physical cause of the destabilizing effects in the superadiabatic regions is revealed
 by Fig.~\ref{work_theo} (middle). 
The Lagrangian perturbation of entropy exhibits a rapid variation 
that occurs mainly in the super-adiabatic layer  and in the atmospheric layers.
 As the frequency of the mode increases, the amplitude of this variation 
 (which is a  spatial oscillation  as seen in the next section) also 
 increases. The wavelenth of this spatial oscillation  decreases with increasing frequency.  This causes a similar
 behavior of the cumulated work.

\subsection{Oscillation of entropy fluctuations}
\label{oscill_s}

To understand the behaviour of $\delta S$ in this region, let us first examine the fluctuations of radiative 
and convective luminosity appearing in the energy equation (\eq{energy}).

\begin{figure}
\begin{center}
\includegraphics[height=6cm,width=9cm]{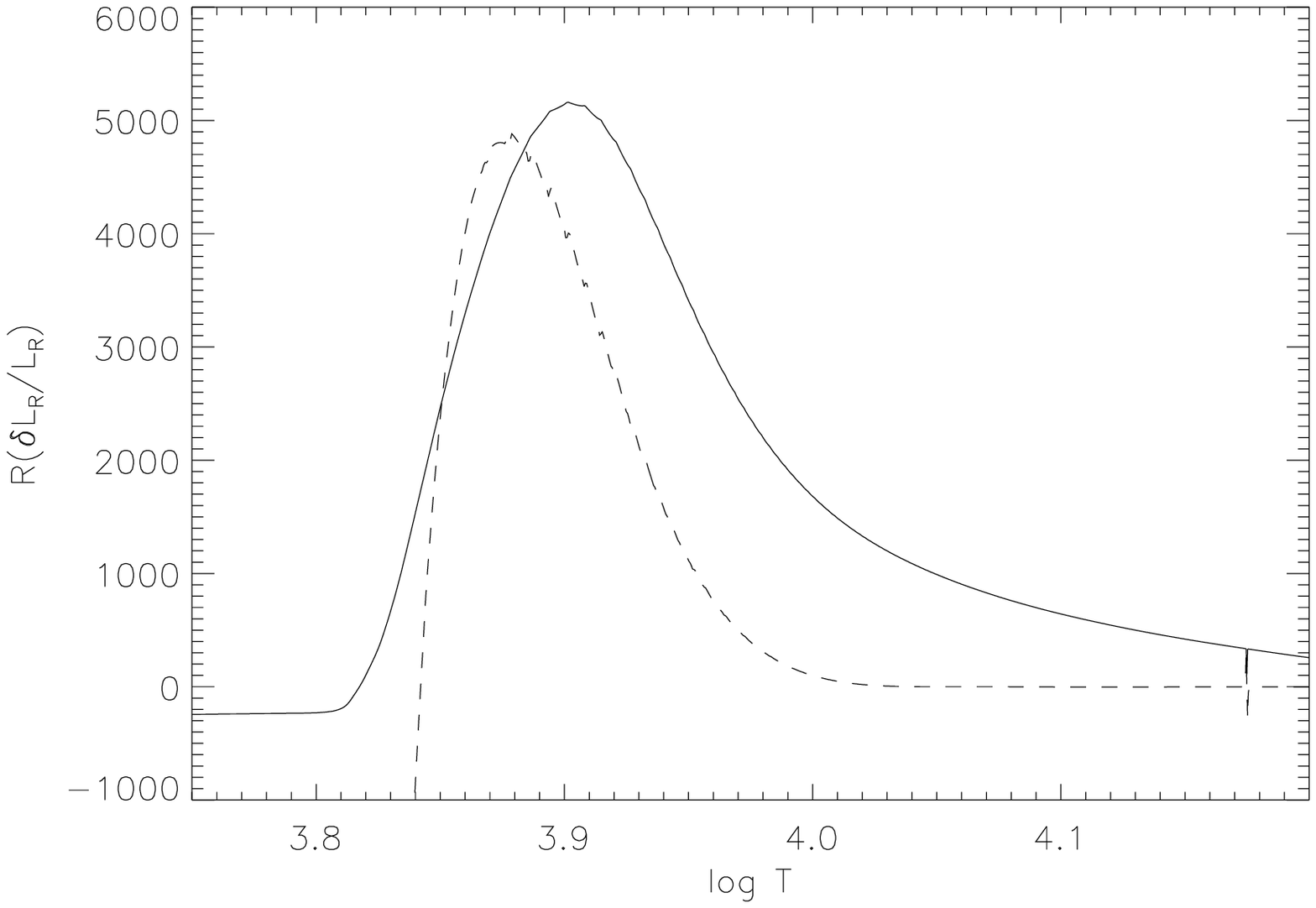}\\
\includegraphics[height=6cm,width=9cm]{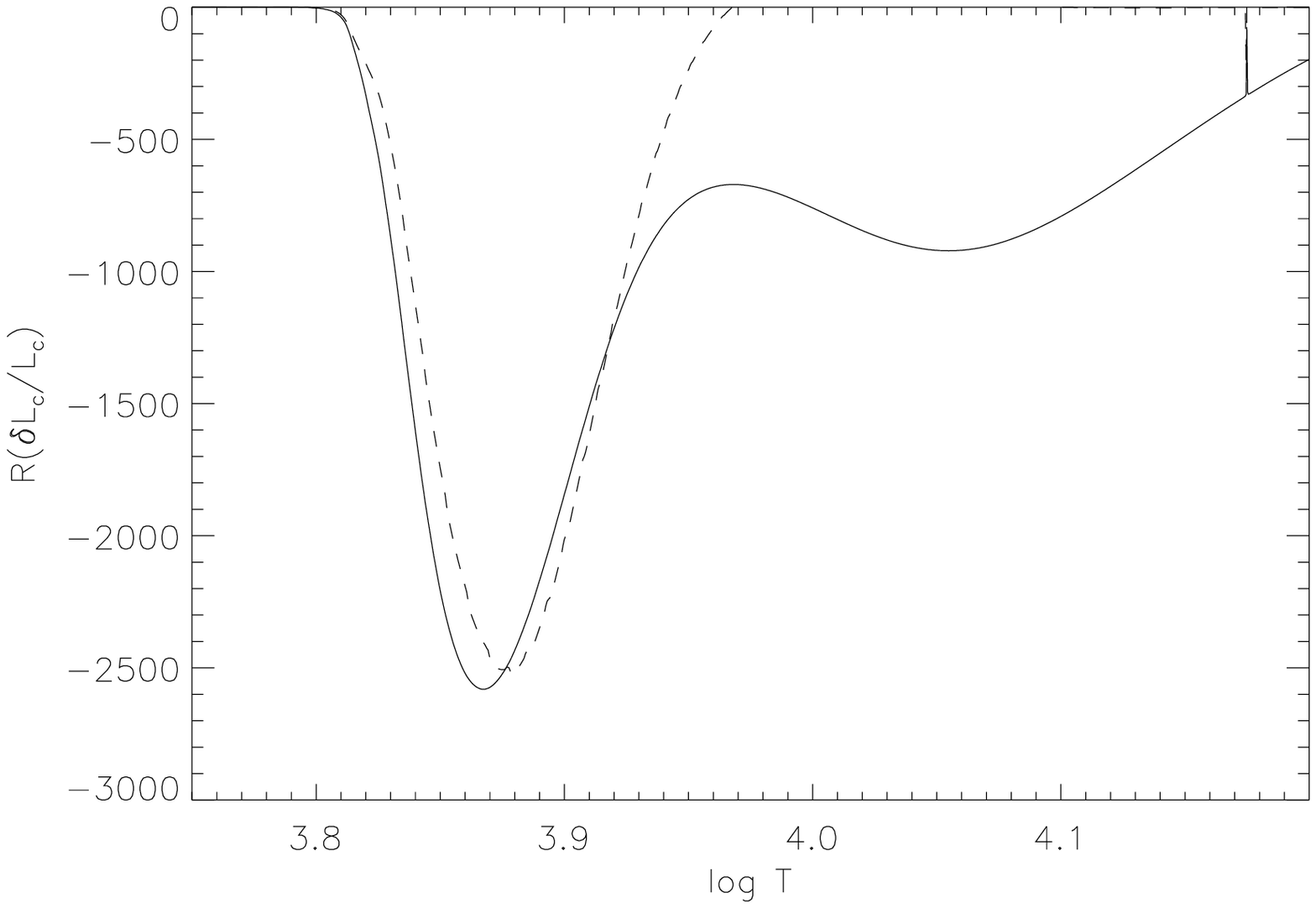}
\caption{{\bf Top:} Real part of the radiative luminosity perturbation for the mode of radial order $n=23$ versus logarithm of temperature. The solid line represents the full non-adiabatic computation as described in Sect.~\ref{compute_damping}, while the dashed line represents the approximate expression given by \eq{deltaLr3}.  The rapid variation near $\log T = 4.17$ is the result of the presence of a node of the eigenfunction.
 {\bf Bottom:} Real part of the convective luminosity perturbation for the mode of radial order $n=23$ versus logarithm of temperature. The solid line represents the full non-adiabatic computation as described in Sect.~\ref{compute_damping}, while the dashed line represents the approximate expression given by \eq{approx_Lc}. 
 }
\label{approx_L}
\end{center}
\end{figure}

In the diffusion approximation, the fluctuations of radiative luminosity reads
\begin{align}
\label{deltaLr1}
\frac{\delta L_{\rm R}}{L_{\rm R}} = 2\frac{\xi_r}{r} + 3 \frac{\delta T}{T} - \frac{\delta \kappa}{\kappa} 
- \frac{\delta \rho}{\rho} + \frac{{\rm d}\delta T / {\rm d}r}{{\rm d}T / {\rm d}r} - \deriv{\xi_r}{r}
\end{align}
where $\xi_r$ is the mode's radial displacement, $\delta T$ the Lagrangian perturbation of temperature, $\delta \kappa$ the perturbation of opacity, and $\kappa$ the opacity. 
By using the perturbed continuity equation \eq{deltaLr1}, becomes, for radial modes, 
\begin{align}
\label{deltaLr2}
\frac{\delta L_{\rm R}}{L_{\rm R}} =  \frac{T}{{\rm d}T / {\rm d}r} \, \deriv{}{r} \left(\frac{\delta T}{T}\right) + 4 \frac{\delta T}{T} - \frac{\delta \kappa}{\kappa}
\end{align}
where we have neglected $\xi_r / r$ compared to $\partial \xi_r / \partial r$. This assumption is valid for radial p modes \citep[see][ for details]{Belkacem08a}. 
We further assume that, in the super-adiabatic region, perturbation of temperature fluctuations and opacity 
are dominated by entropy fluctuations, so that
\begin{align}
\label{non-ad}
\frac{\delta T}{T} \backsim \frac{\delta S}{c_v} \quad {\rm and} \quad 
\frac{\delta \kappa}{\kappa} \backsim \kappa_T \frac{\delta S}{c_v}
\end{align}
where $c_v = \left(\partial U / \partial T \right)_\rho$ with $U$ the internal energy per unit mass, and $\kappa_T = \left(\partial \ln \kappa / \partial \ln T \right)_\rho$. 
Hence, inserting Eqs.~(\ref{non-ad}) in \eq{deltaLr1} we obtain
\begin{align}
\label{deltaLr3}
\frac{\delta L_{\rm R}}{L_{\rm R}} \backsim \frac{T}{{\rm d}T / {\rm d}r} \, \deriv{}{r} \left( \frac{\delta S}{c_v} \right) + (4-\kappa_T) \frac{\delta S}{c_v}
\end{align}
The approximate expression \eq{deltaLr3}, even if imperfect, captures the main behaviour 
of $\delta L_R / L_R$ in the superadiabatic boundary region, as shown by Fig.~\ref{approx_L} (top). 
Note that the disagreement observed in the inner layers in Fig.~\ref{approx_L} (top panel) is due to 
the approximation \eq{non-ad} since for those layers the density fluctuations are dominant. However, 
we are mainly interested in the super-adiabatic region ($\log T < 3.9$) where \eq{deltaLr3} is sufficiently valid 
for our purpose. 

We now turn to the Lagrangian perturbation of convective luminosity. It is dependent on the adopted time-dependent 
treatment of convection. Consistent with Sect.~\ref{compute_damping}, we use the formalism developed by \cite{MAD05}. A good approximation of their Eq.~(18) in the super-adiabatic layer, as shown by Fig.~\ref{approx_L} (bottom), is 
\begin{align}
\label{approx_Lc}
\frac{\delta L_{\rm c}}{L_{\rm c}} & \backsim \psi \deriv{\delta S}{S} 
= \psi \left[ \deriv{}{r} \left( \frac{\delta S}{c_v} \right)  + \deriv{\ln c_v}{r} \left(\frac{\delta S}{c_v}\right)\right] \frac{T}{{\rm d} T / {\rm d}r } \, ,
\end{align}
with  
\begin{align}
\label{def_psi}
\psi \backsim 
C \left(1+ \frac{\left( (i+\beta)\sigma \tau_c + 2 \omega_R \tau_c+1 \right)D}{B+\left((i+\beta)\sigma \tau_c + 1\right)D}\right) \, .  
\end{align}
where $\tau_c$ is the life-time of convective elements, $\omega_R$ is the characteristic cooling frequency of the turbulent eddies, and 
\begin{align}
B &= \frac{i\sigma \tau_c+\Lambda}{\Lambda} \, , \\
C &= \frac{\omega_R \tau_c+1}{(i+\beta) \sigma \tau_c + \omega_R \tau_c+1} \, ,\\
D &= \frac{C}{\left(\omega_R \tau_c+1\right)} \, .
\end{align}
where $\Lambda=8/3$ is a constant introduced by \cite{Unno67} to close the equation of motion
describing convection, and $\omega_R$ is the characteristic cooling frequency of turbulent eddies due to
radiative losses \citep[see Eq. (C12) of][]{MAD05}. 

 We are now interested in obtaining the equation that qualitatively explains the oscillation observed in Fig.~\ref{work_theo} (middle panel). Hence, one has to exhibit in an analytical way the frequency dependence of the entropy fluctuations ($\delta S$). To this end, we will use two different assumptions. The first, and most immediate way is to assume that in the energy equation (\eq{energy}) ${\rm d} \delta L / {\rm d}r \backsim  \delta L / H_p$. This is a crude approximation, but which permits to immediately exhibit the role of the $\mathcal{Q}$ factor. Then 
using \eq{deltaLr3} and \eq{approx_Lc}, one obtains
\begin{align}
\label{equation_osc}
\deriv{}{\ln T} \left( \frac{\delta S}{c_v} \right)+ \lambda \left( \frac{\delta S}{c_v} \right) = 0 \,, \quad {\rm with} \quad 
\lambda = \mathcal{A} - i \, \mathcal{B}\, , 
\end{align}
where 
$\mathcal{A}$ and $\mathcal{B}$ are defined by 
\begin{align}
\label{lambda}
\mathcal{A} &=  \left(\frac{L_c}{L} \psi \deriv{\ln c_v}{\ln T} + \frac{L_R}{L} (4-\kappa_T) \right)\left( 1 + (\psi-1) \frac{L_c}{L} \right)^{-1}  \nonumber \\
\mathcal{B} &= \mathcal{Q} \left[ 1 + (\psi-1) \frac{L_c}{L} \right]^{-1}  \, ,  
\end{align} 
where we have defined the ratio $\mathcal{Q}$ as
\begin{align}
\label{def_Q}
\mathcal{Q} = \omega \tau \, , \quad \mbox{with} \quad \tau^{-1} = \frac{L}{4\pi r^2\rho c_v T H_p}
\end{align}
with $\tau$ is a local thermal time-scale. Note that we have neglected the imaginary part of $\sigma$ in \eq{def_Q}. We stress that this thermal time-scale  can be recast into
\begin{align}
\tau^{-1} = \tau_{\rm conv}^{-1} + \tau_{\rm rad}^{-1}
\end{align}
where $ \tau_{\rm conv}$ and $ \tau_{\rm rad}$ are associated with the convective and radiative luminosities, respectively.
From \eq{equation_osc}, the oscillatory part of the final solution is 
$\left( \delta S / c_v \right) \propto \exp \left[- i \, \int \mathcal{B} \,  {\rm d}\ln T \right]$, 
which explains the oscillatory behaviour of entropy perturbations in the super-adiabatic  layers and its frequency dependence. 

An alternative way to proceed is to use  
the energy equation (\eq{energy}) 
together with \eq{deltaLr3} and \eq{approx_Lc}, one obtains the second order differential equation
\begin{align}
\label{eq_osc_S_second_order}
\mathcal{F} \deriv{^2}{\ln T^2} \left(\frac{\delta S}{c_v}\right) + \mathcal{G} \deriv{}{\ln T} \left(\frac{\delta S}{c_v}\right) + \mathcal{H} \left(\frac{\delta S}{c_v}\right) = 0
\end{align}
where 
\begin{align}
\mathcal{F}  &= 1 + (\psi-1) \frac{L_c}{L} \nonumber \\
\mathcal{K}  &=\frac{L_R}{L} (4-\kappa_T)  + \psi \frac{L_c}{L} \deriv{\ln c_v}{\ln T} \\
\mathcal{G} &= \deriv{\mathcal{F}} {\ln T} + \mathcal{K}   \nonumber \\
\mathcal{H} &= \deriv{\mathcal{K} }{\ln T} - i \mathcal{Q} \left(\frac{H_p}{H_T}\right)
\nonumber 
\end{align}
where $H_T$ is the temperature scale-height. 

 To derive an analytical solution of \eq{eq_osc_S_second_order} is not trivial. Hence, further simplifications are needed. We then assume the coefficients $\mathcal{F}, \mathcal{G}, \mathcal{H}$ are constant. 
Assuming solutions of the form $\left( \delta S / c_v \right) \propto e^{k \ln T}$, one has the solutions for $k$
\begin{align}
\label{wavenumber}
k_{1,2} = \frac{-\mathcal{G} \pm \left[ \mathcal{G}^2 - 4\mathcal{H}\mathcal{F}  \right]^{1/2}}{2 \mathcal{F}} 
\end{align}
At the maximum of the super-adiabatic gradient, the radiative luminosity dominates over the convective ones. Hence, we further neglect the ratio $L_c/L$ compared with $L_R/L$. \eq{wavenumber} then simplifies to 
\begin{align}
\label{wavenumber2}
k_{1,2} = -\frac{1}{2} \left[ \mathcal{G} \pm \left( \mathcal{G}^2 - 4\deriv{G}{\ln T} + 4i \mathcal{Q} \left(\frac{H_p}{H_T}\right)  \right)^{1/2} \right] 
\end{align} 
From \eq{wavenumber2}, one concludes that for $\mathcal{Q} \ll 1$, $k$ is real and $\delta S$ does not oscillate. This corresponds to the limit of low-frequency modes for which both $k$ and the imaginary part of $\delta S$ are small, as confirmed by the full numerical computation presented in Fig.~\ref{work_theo} (middle panel). 
In contrast, for $\mathcal{Q}\gg 1$ (i.e., for large frequencies) the imaginary part of the wavenumber increases as depicted by  Fig.~\ref{work_theo} (middle panel). 

Eventually, both methods to derive the frequency behaviour of $\delta S$ converge toward the same conclusion, i.e. that the factor $\mathcal{Q}$ explains the oscillation of entropy fluctuations and its frequency dependence. 

\end{document}